\documentclass[12pt,aps]{revtex4}
\usepackage{epsfig,amsmath,amssymb}

\newcommand{\eq}{{\, \equiv\, }}
\newcommand{\fr}[1]{
             \frac{#1}}
\newcommand{\bea}{\begin{eqnarray}}
\newcommand{\eea}{\end{eqnarray}}

\newcommand{\chibar}{\overline{\chi}}

\newcommand{\ket}{{\rangle }}

\newcommand{\bra}{{\langle }}
\newcommand{\gc}{\bra\fr{\alpha_s}{\pi}G^2\ket}

\newcommand{\ga}{{g_{{\mathcal A}}}}

\begin{document}
\title{On the    Decay Modes   $\overline{B_{d,s}^0} \rightarrow \, \gamma D^{*0}$} 

\vspace{0.5cm}

\author{J.A. Macdonald S\o rensen}
\email{j.a.m.d.sorensen@fys.uio.no}
\affiliation{Department of Physics, University of Oslo,
P.O.Box 1048 Blindern, N-0316 Oslo, Norway}
\author{J.O. Eeg}
\email{j.o.eeg@fys.uio.no}
\affiliation{Department of Physics, University of Oslo,
P.O.Box 1048 Blindern, N-0316 Oslo, Norway}

\vspace{0.5 cm}

\begin{abstract}

The various decay modes of the type 
$B \rightarrow \gamma \, D^* $ are dynamically different. 
In general there are factorizable contributions, and there are pole diagrams 
 and pseudoscalar exchange contributions  at meson level.
The purpose of this paper is to point out that the decay modes 
  $\overline{B_{d,s}^0} \rightarrow \, \gamma D^{*0} $
have  negligible  contributions from such mechanisms,
in contrast to the decay modes  $\overline{B_{d,s}^0} \rightarrow \,
\gamma \overline{D^{*0}} $
and  $B^- \rightarrow \, \gamma D_{s,d}^{*-} $.
However, for the decay modes $\overline{B_{d,s}^0} \rightarrow \,
\gamma D^{*0}$
there are  non-factorizable contributions due to emision of
soft gluons, and such non-factorizable contributions are found to
dominate the amplitudes for these latter decay modes.

We estimate the branching ratio for these modes 
 in the heavy quark limits, both for  the $b$- and the $c$-
quarks, and obtain a value $ \simeq \, 1.6 \times 10^{-6}$ for 
$\overline{B_{d}^0} \rightarrow \, \gamma D^{*0}$, and 
 $ \simeq \, 8 \times 10^{-7}$ for
 $\overline{B_{s}^0} \rightarrow \, \gamma D^{*0}$.
We expect large corrections to this limit because the energy gap
between the  $b$- and $c$- quark masses are significantly bigger than  1 GeV.
However, we expect that our estimate for 
 $\overline{B_{d,s}^0} \rightarrow \, \gamma D^{*0}$
 gives the right order of magnitude for   the {\em amplitudes}.

\end{abstract}

\maketitle

\vspace{1cm}



\newpage

\section{Introduction}
There is presently great interest in decays of $B$-mesons, 
due to numerous experimental results coming from BaBar and Belle. 
Later  LHC will provide data for such processes. 
$B$-decays of the type $B \rightarrow \pi \pi$ and $B \rightarrow K
\pi$,  
where  the energy
 release is big  compared to the light
meson masses,  has been treated within
{\it QCD factorization} and {\em soft collinear effective  theory} (SCET)
 \cite{BBNS}. In these cases  
 the amplitudes factorize into  products
of two matrix elements of weak currents in the high energy limit, 
and  non-factorizable corrections of order $\alpha_s$
  can be calculated perturbatively.

 The decays $B \rightarrow \pi \pi,  K \pi$ are typical heavy to light decays.  
It was pointed out in previous papers \cite{EFHP} that 
 for various decays of the type $\bar{B} \rightarrow D \bar{D}$, which
 are of heavy to heavy type,
  the methods of \cite{BBNS} are not expected to hold because  the energy release
is of order 1 GeV.
(Here $\bar{B}$, $D$, and $\bar{D}$ contain a heavy $b$, $c$, and
anti-$c$ quark respectively).
In this paper we consider  decay modes of the type 
$B \rightarrow  \gamma \, D^* $. Such modes have been studied in the 
literature \cite{BDg,GriLe} for some time. We restrict ourselves to
processes  where the $b$-quark decays. This means  
 the quark level  processes  
$b \bar{q} \rightarrow \gamma c \bar{u}$,
$b \bar{q} \rightarrow  \gamma u \bar{c}$, and 
$b \bar{u} \rightarrow \gamma c \bar{q}$, where $q=d$ or  $q=s$.
Processes where the anti- $b$-quark decays proceed analogously.

 Formally,  decays of the type  $\bar{B} \rightarrow
\gamma \, D$ is a heavy to heavy transition in the heavy quark 
limits $(1/m_b) \rightarrow 0$ {\em and } $(1/m_c) \rightarrow 0$, 
and in ref. \cite{GriLe} the  decay of a charged $B$-meson was studied
 within {\em heavy quark effective theory} (HQEFT) \cite{neu} and
{\em heavy light chiral perturbation theory} (HL$\chi$PT)
\cite{itchpt}.
This framework was also used to study the  Isgur-Wise
function for the $B \rightarrow D$ transition currents,
which is also  a heavy to heavy transition where
chiral loops (in terms of HL$\chi$PT) and 
 $1/m_{b,c}$ corrections (in terms of HQEFT) have been added  \cite{IW}.
In the present paper we will also stick to this framework, although it is not
expected to hold for precise numerical estimates
because the enery gap between the $b$- and the $c$-scale is
substantial, 
namely about three times the chiral symmetry breaking scale.

First,  decay modes of the type 
$B \rightarrow  \gamma \, D^* $
might have  substantial factorizable contributions,
of pole or non-pole type. These pole diagrams are present  only for
radiation from charged $B$- or $D$-mesons. Second,  there are also
 meson exchange contributions. These will be  chiral loop contributions
 in the HQEFT limits (for the $b$ and $c$-quarks).
Such meson exchange diagrams, which are non-factorizable and $1/N_c$
suppressed, are present for the decay modes
  $\overline{B_{d,s}^0} \rightarrow \,
\gamma \overline{D^{*0}} $
and  $B^- \rightarrow \, \gamma D_{s,d}^{*-} $.

The purpose of this paper is to point out that the decay modes
 $\overline{B_{d,s}^0} \rightarrow \, \gamma D^{*0}$
have almost zero contribution from the factorizable and the 
meson exchange mechanisms. However,
 these decay modes have significant contributions from
  soft gluon emision.
Such  non-factorizable (colour suppressed $\sim 1/N_c$) contributions
to $B - \bar{B}$ mixing \cite{ahjoeB}, $B \rightarrow D \bar{D}$ \cite{EFHP}
 and $B \rightarrow D \eta'$ \cite{EHP}
 decays  are 
calculated in terms of the (lowest dimension) gluon condensate within
a recently developed {\em heavy light chiral quark model} (HL$\chi$QM) \cite{ahjoe},
which is based on the HQEFT \cite{neu}. They have also been studied in
 the  light sector for  $K - \bar{K}$ mixing and $K \rightarrow 2
 \pi$ decays \cite{BEF}.
We estimate the branching ratios for 
 $\overline{B_{d,s}^0} \rightarrow \, \gamma D^{*0}$
 in the heavy $b$- and $c$-quark limits.
Note that the decay modes  $\overline{B^0_{d,s}} \rightarrow \gamma \, D^{*0}$
  and $\overline{B^0_{d,s}} \rightarrow \gamma \, \overline{D^{*0}}$ proceed
 differently. In
the last case there are substantial meson exchange contributions.

In the next section (II) we present the
 weak four quark Lagrangian and its factorizable and non-factorizable
 matrix elements.
In section III we  present the framework of HQEFT and HL$\chi$PT, and
in section IV we  calculate the 
non-factorizable
matrix elements due to soft gluons expressed through the (model
dependent) quark condensate. 
 In section V we give the results and 
conclusion.

\section{The   weak quark Lagrangian and its matrix elements}

\begin{figure}[t]
\begin{center}
   \epsfig{file=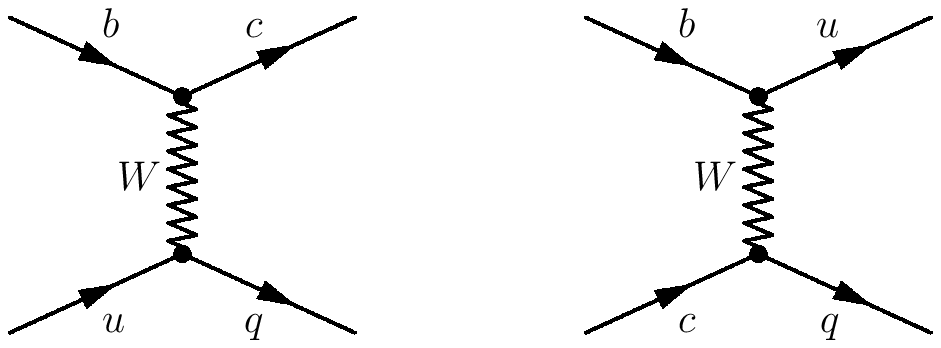,width=12cm}
\caption{Tree level W-exchange leading to the effective Lagrangian in
eq. (\ref{Lquark}). The left diagram 1a) gives rise to $Q_{1,2}$, and
the right diagram 1b) gives rise to $Q_{3,4}$.}
 \label{fig:TreeLev}
\end{center}
\end{figure}

Based on the electroweak and quantum chromodynamical interactions,
one constructs an effective non-leptonic Lagrangian at
quark level in the standard way: 
 \begin{equation}
 {\mathcal L}_{W}=  \sum_i  C_i(\mu) \; Q_i (\mu) \; ,
 \label{Lquark}
\end{equation}
where all information of the short distance (SD) loop 
effects above a renormalization scale $\mu$
is contained in the Wilson coefficients $C_i$.
 In our case there are four relevant
operators 
\begin{eqnarray}
Q_{1}  = 4  (\overline{q}_L \gamma^\alpha  b_L )  \; 
           ( \overline{c}_L \gamma_\alpha  u_L )  
\qquad  ,  \; \; \; \;
Q_{2}  =  4 \,  ( \overline{c}_L \gamma^\alpha  b_L ) \;  
           ( \overline{q}_L \gamma_\alpha  u_L ) \; , \\
\label{Q12} 
Q_{3}  = 4  (\overline{q}_L \gamma^\alpha  b_L )  \; 
           ( \overline{u}_L \gamma_\alpha  c_L )  
\qquad  ,  \; \; \; \;
Q_{4}  =  4 \,  ( \overline{u}_L \gamma^\alpha  b_L ) \;  
           ( \overline{q}_L \gamma_\alpha  c_L ) \; ,
\label{Q34} 
\end{eqnarray} 
for $q = d,s$. 
This effective Lagrangian is based on the interactions in Fig.~\ref{fig:TreeLev}
and hard gluon corrections to these diagrams.
Operators from penguin diagrams may also contribute, but have small Wilson coefficients.
The coefficients $C_{1,2}$ and  $C_{3,4}$ have different KM structures.
We may write
\begin{eqnarray} 
 C_{i} = - \frac{G_F}{\sqrt{2}} (V_{cb}V_{uq}^*) \, a_{i} \quad ;
 \quad 
C_{j} = - \frac{G_F}{\sqrt{2}} (V_{ub}V_{cq}^*) \,  a_{j} \; ,
\label{Wilson} 
\end{eqnarray} 
for $i=1,2$ and $j=3,4$ respectively. 
Here the ``reduced'' Wilson coefficients $a_i$  (for  $i=1,2,3,4$) are
dimensionless numbers. Furthermore,  in terms
of the Wolfenstein parameter $\lambda$, we have 
$V_{cb}V_{ud}^* \sim {\cal O}(\lambda^2)$. For $q=s$ the KM factors 
$V_{cb}V_{us}^*$ and $V_{ub}V_{cs}^*$ are both $\sim {\cal  O}(\lambda^3)$,
 while   $V_{ub}V_{cd}^* \sim  {\cal O}(\lambda^4)$.
 At the scale
$\mu = M_W$, when perturbative QCD is switched off, one has
 $a_{1,3}=0$ and  $a_{2,4}=1$. 
 At the scale $\mu =
m_b$,  $a_{1,3} \sim 10^{-1}$ and negative,  and $a_{2,4}$ are still $
\sim 1$ \cite{BuBuLa}. (In practice, $a_1=a_3$ and $a_2=a_4$.)
 Extrapolating the Wilson coefficients (naively) down to 
 $\mu \sim \Lambda_\chi \sim$ 1 GeV, which is the matching scale
between short and long distance effects within our
 framework \cite{ahjoeB,EFHP,EHP},
  we obtain $a_{2,4} \simeq 1.17$ and 
$a_{1,3} \simeq -0.37$ \cite{EHP}. Alternatively, one might perform perturbative QCD
within HQEFT as done in  \cite{GKMWF} and used
 in \cite{EFHP} for $B \rightarrow D \bar{D}$, but numerical
 differences will be small.

One may also think of operators like
\begin{eqnarray}
 e  F_{\mu \nu}  (\overline{q}_L \gamma^\mu  b_L )  \; 
           ( \overline{c}_L \gamma^\nu  u_L )  \quad , \qquad 
 e  F_{\mu \nu}  (\overline{q}_L \sigma^{\alpha \beta} F_{\alpha \beta}
 \gamma^\mu  b_L )  \; 
           ( \overline{c}_L \gamma_\nu  u_L ) \; .
\label{QS} 
\end{eqnarray} 
 However, such operators are of dimension eight, and dominated at
 low momenta which make a short distance treatment dubious.

In the {\it factorized} limit (-no strong interactions between the two
quark currents) we obtain  
the amplitude for  $\overline{B_q^0} \rightarrow \gamma \, D^{*0} $ 
obtained from (\ref{Lquark}) and (\ref{Q12}): 
\begin{eqnarray}
\langle \gamma \, D^{*0} | {\mathcal L}_W| \overline{B_q^0} \rangle_F \,  = \,
 \; &4& \left(C_1 + \frac{C_2}{N_c}\right) \, \Big( 
\langle D^{*0}| \overline{c_L}\gamma_\mu  u_L |0 \rangle
\langle \gamma |\overline{q_L}\gamma_\mu b_L|\overline{B^0} \rangle \;
 \Big.   \nonumber \\
\, &+&  \Big. 
\langle \gamma \, D^{*0}| \overline{c_L}\gamma_\mu  u_L |0 \rangle
\langle 0 |\overline{q_L}\gamma_\mu b_L|\overline{B^0} \rangle \; \Big) ,
\label{FactorizedP1} 
\end{eqnarray}
where  the subscript $F$ means ``factorized''.
For  $\overline{B_q^0} \rightarrow \gamma \,  \overline{D^{*0}}$
 we obtain the same expression with $C_{1,2}$ replaced by  $C_{3,4}$
and with $c$ and  $u$ interchanged.
Thus, the neutral decays have negligible factorized contributions
proportional to $a_{nf} =(a_{1,3}+a_{2,4}/N_c)$, which is of order $10^{-2}$.

For  the charged case $B^- \rightarrow \gamma \, D_{q}^{*-} \,$ we obtain:
\begin{eqnarray}
\langle \gamma \, D_q^{*-} | {\mathcal L}_W| B^- \rangle_F \,  = \,
 \; &4& \left(C_4 + \frac{C_3}{N_c}\right) \, \Big( 
\langle D_q^{*-}| \overline{q_L}\gamma_\mu  c_L |0 \rangle
\langle \gamma |\overline{u_L}\gamma_\mu b_L| B^- \rangle \;
 \Big.   \nonumber \\
\, &+&  \Big.
\langle D_q^{*-} \gamma| \overline{q_L}\gamma_\mu  c_L |0 \rangle
\langle 0 |\overline{u_L}\gamma_\mu b_L| B^- \rangle \; \Big) \; .
\label{FactorizedP3} 
\end{eqnarray}
\begin{figure}[t]
\begin{center}
   \epsfig{file=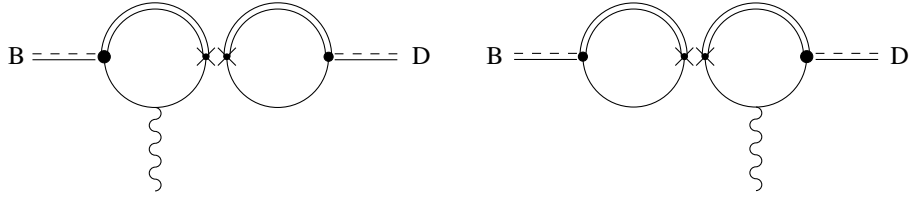,width=12cm}
\caption{Factorized contributions for $B  \rightarrow \gamma D^*$. 
 The combined  dashed and full  lines represent heavy mesons, the double lines
 represent heavy quarks, and the single lines light quarks The wavy
 line is a photon. a) Emision of photon from $B$-meson. 
b) Emision of photon from $D$-meson }
\label{fig:BDfact}
\end{center}
\end{figure}
Thus the charged decays have substantial factorizable contributions
proportional to $a_f = \left(a_{4} + a_{3}/N_c\right) \sim 1$.

In order to study non-factorizable contributions at quark level, 
we  use the 
following relation between the generators of $SU(3)_c$ ($i,j,l,n$
are colour indices running from 1 to 3):
\begin{equation}
\delta_{i j}\delta_{l n}  =   \fr{1}{N_c} \delta_{i n} \delta_{l j}
 \; +  \; 2 \; t_{i n}^a \; t_{l j}^a \; ,
\label{fierz}
\end{equation}
where $a$ is the color octet  index.
Then the operators $Q_{1,2}$ may, by means of a Fierz transformation,
be  written in the following way :
\begin{eqnarray}
Q_{1,3}  =  \frac{1}{N_c} Q_{2,4} + 2 \widetilde{Q}_{2,4}
\qquad , \; \; \;    \,Q_{2,4} =  \frac{1}{N_c} Q_{1,3} + 2 \widetilde{Q}_{1,3}
\; \;  , 
\label{QFierz} 
\end{eqnarray}
where  the  operators with the ``tilde'' contain colour matrices:
\begin{eqnarray}
\widetilde{Q_{1}}  = 4  (\overline{q}_L \gamma^\alpha t^a  b_L )  \; \,
           ( \overline{c}_L \gamma_\alpha t^a u_L ) \,  
\qquad , \; \;  \;
\widetilde{Q_{2}}  =  4 \,  ( \overline{c}_L \gamma^\alpha t^a b_L )  \; \,
           ( \overline{q}_L \gamma_\alpha t^a u_L ) \; .\label{QCol12} \\ 
\widetilde{Q_{3}}  = 4  (\overline{q}_L \gamma^\alpha t^a  b_L )  \; \,
           ( \overline{u}_L \gamma_\alpha t^a c_L ) \,  
\qquad , \; \;  \;
\widetilde{Q_{4}}  =  4 \,  ( \overline{u}_L \gamma^\alpha t^a b_L )  \; \,
           ( \overline{q}_L \gamma_\alpha t^a c_L ) \; .
\label{QCol34} 
\end{eqnarray}  

To obtain
 physical amplitudes, one has to calculate the hadronic matrix elements
of the quark operators $Q_i$ and  $\widetilde{Q_i}$ within some  framework
describing  long distance (LD) effects.

The non-factorizable amplitude for 
$\overline{B_q^0} \rightarrow \, \gamma \,  D^{*0}$, 
with one gluon emision obtained from
the coloured operators in (\ref{QCol12}) and  (\ref{QCol34})
 might be written in a quashi factorizable way interms of octet
 gluonic intermediate state:
\begin{eqnarray}
\langle \gamma D^{*0} &|& {\mathcal L}_W| \overline{B_q^0} \rangle_{NFG} \, = \,
8 \, C_2 \, \langle \gamma \, D^{*0}| \widetilde{Q_1}| \overline{B_q^0} \rangle
\nonumber \\
\; &=& \; 8 \, C_2 \, \Big( 
\langle D^{*0}| \overline{c_L}\gamma_\mu  t^a \, u_L |G \rangle
\langle G \gamma |\overline{q_L}\gamma_\mu t^a b_L|\overline{B_q^0} \rangle \;
 \Big.   \nonumber \\
\, &+& 
 \Big. \; 
\langle \gamma D^{*0} | \overline{c_L}\gamma_\mu  t^a u_L |G \rangle
\langle G |\overline{q_L}\gamma_\mu t^a b_L|\overline{B_q^0} \rangle \;
\Big) \; ,
\label{NFactorizedP1} 
\end{eqnarray}

\begin{figure}[t]
\begin{center}
   \epsfig{file=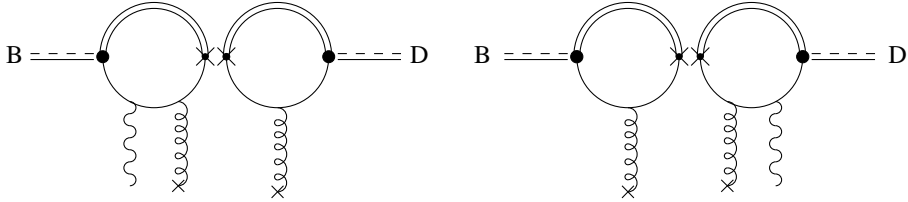,width=12cm}
\caption{Non-factorizable  contributions to $B \rightarrow \gamma D$
from the coloured operators $\widetilde{Q_i}$ within a
quashi-factorizable approximation.The curly lines represent soft gluon
emision ending in vacuum to make a gluon condensate. a) With
additional photon emision from the $B$-meson. b) With additional
photon emision from $D$-meson }
 \label{fig:Nonfact}
\end{center}
\end{figure}

This amplitude is  visualized later in 
figure~\ref{fig:Nonfact}, and may be calculated   within the HL$\chi$QM.
The non-factorizable amplitude for 
$\overline{B_q^0} \rightarrow \gamma \,  \overline{D^{*0}}$ 
with one gluon emision obtained from
the coloured operators
is the same as above with $C_2 \rightarrow C_4$ and with $u$ and $c$-quarks interchanged.
The non-factorizable amplitude for 
$B^- \rightarrow \gamma \,  D_q^{*-}$ with  gluon emision  from
the coloured operators is proportional to $C_3$ and therefore
relatively small.

We observe the following generic pattern: 
 Some  decay modes have
 substantial factorizable contributions proportional to the favorable
 Wilson coefficient linear combination 
 $a_f \equiv (a_{2,4} + a_{1,3}/N_c)$, which is close to one. In this case there
 are contributions from the coloured operators  $\widetilde{Q}_{1,3}$
 proportional to $2 a_{1,3}$ of moderate importance.
For other modes there might be 
 factorizable matrix elements   proportional to the non-favorable
coefficient 
$a_{nf} \equiv (a_{1,3} +  a_{2,4}/N_c)$ which is close
 to zero (of order $10^{-2}$ or smaller)
 at our matching scale $\mu = \Lambda_\chi$.
 In these cases there are
 substantial contributions proportional to $2 a_{2,4}$ from the coloured
 operators  $\widetilde{Q}_{1,3}$.

In terms of the $B$-meson field $\Phi$, the $D^*$-meson field $V^\mu$,
and the electromagnetic field tensor $F_{\mu \nu}$, we can write down the effective
Lagrangian to first order in the photon momentum, consistent with the
heavy quark  limits:
\bea
{\cal L}_{eff} \, = \, 
 A^{(+)} \, i \epsilon_{\mu \nu \alpha \beta} \,\Phi \,  F^{\mu \nu} \, V^\alpha
 v_b^\beta \; + \;  A^{(-)} \; \, \Phi \,  F_{\mu \nu} \, V^\mu \, v_b^\nu\ \; ,
\label{LeffBD}
\eea 
where the positive and negative parity amplitudes $A^{(\pm)}$
 depend on hadronic parameters, and the meson masses $M_{B,D}$.

\section{Theoretical framework for the heavy quark limits}

Our calculations will be based  on 
 HQEFT \cite{neu}, which is a systematic $1/m_Q$ expansion in the
 heavy quark mass $m_Q$.
 The heavy quark fields $Q (=b, c$ 
 $\overline{c}$)
 are replaced with a ``reduced''
field $Q_v^{(+)}$ for a heavy quark, and $Q_v^{(-)}$ for a heavy
antiquark (in the case  $\overline{c}$). 
 The Lagrangian for heavy
quarks is:
\begin{equation}
{\mathcal L}_{HQEFT} =   \pm \overline{Q_v^{(\pm)}} \, i v \cdot D \, Q_v^{(\pm)} 
 + {\mathcal O}(m_Q^{- 1}) \; ,
\label{LHQEFT}
\end{equation}
where $v$ is the velocity of the heavy quark, 
and $D_\mu$ is the covariant derivative containing the gluon and
the photon fields.
In \cite{ahjoeB} the  $1/m_Q$ corrections were calculated for $B
 -\overline{B}$  -mixing. In this paper these will not be considered.

Integrating out the heavy and light quarks, the effective Lagrangian up
to ${\mathcal O}(m_Q^{-1})$ can be written as 
\cite{itchpt,ahjoe}
\begin{equation}
{\mathcal L} =  \mp Tr\left[\overline{H^{(\pm)}_{a}}
i v \cdot  D_{ba}
H^{(\pm)}_{b}\right]\, -\, 
\ga \, Tr\left[\overline{H^{(\pm)}_{a}}H^{(\pm)}_{b}
\gamma_\mu\gamma_5 {\mathcal A}^\mu_{ba}\right]\, + \, ...
,\label{LS1}
\end{equation}
where the ellipses denote terms not relevant in this paper.The indices
$a,b=1,2,3$ are indices corresponding to the quark flavours $u,d,s$
and  $H_a^{(\pm)}$ is the heavy meson field  containing
 a spin zero and spin one boson and ${\mathcal A}^\mu$ is an axial field:
\begin{eqnarray}
&H_a^{(\pm)} & \eq  P_{\pm} (P_{a \mu}^{(\pm)} \gamma^\mu -     
i P_{a 5}^{(\pm)} \gamma_5)  \; \; ; \; 
{\mathcal A}_\mu\eq  -  \fr{i}{2}
(\xi^\dagger\partial_\mu\xi
-\xi\partial_\mu\xi^\dagger) \;  , 
\label{HAV}
\end{eqnarray}
where $P_\pm$ are projecting operators $P_\pm=(1 \pm \gamma \cdot
 v)/2$, and  $v$ is the velocity of the heavy quark. here
 $\xi\equiv exp{(i \Pi/f)}\,$, 
where $f$ is the bare pion coupling, and $\Pi$ is a  3 by 3 matrix
which contains the Goldstone bosons $\pi,K,\eta$ in the standard way.
The axial chiral coupling is 
$\ga \simeq 0.6$. Eqs. (\ref{LS1}) and  (\ref{HAV}) are 
used for the chiral loop contributions within HL$\chi$PT.
 The covariant derivative is given by 
$i D^\mu_{ba} = i \delta_{ba} \partial_\mu - \, e \widetilde{Q_\xi} \,
A^\mu$ , where $\widetilde{Q_\xi} = \xi Q \xi^\dagger R + \xi^\dagger Q
\xi L$ and  $A^\mu$ is the photon field.

The simplest  way to calculate the matrix element of  four quark 
operators like $Q_{1-4}$ in eq. (\ref{Lquark}) is by inserting vacuum
 states between the two currents, as indicated in section II.
This vacuum insertion approach (VSA)
corresponds to bosonizing the two currents in $Q_{1-4}$ separately
 and multiply them, i.e. the factorized case.
Based on the symmetry of HQEFT,
 the bosonized current for decay of the $b \bar{q}$
system is \cite{itchpt,ahjoe}:
\begin{equation}
 \overline{q_L} \,\gamma^\mu\, Q_{b v}^{(+)} \;  \longrightarrow \;
    \fr{\alpha_H}{2} Tr\left[\xi^{\dagger} \gamma^\alpha
L \,  H_{b q}^{(+)} \right]
  \; ,
\label{J(0)}
\end{equation}
where $Q_{b v}^{(+)}$ is a heavy $b$-quark field, $v=v_b$ is its velocity, and
$H_{b q}^{(+)}$ is the corresponding heavy meson field for $\overline{B_q}$.
This bosonization has to be compared with the 
 matrix elements defining  the meson decay 
constants $f_H \, (H=B,D)$.  Before  QCD for scales $\mu < m_Q$
 and chiral loop corrections, one has 
$\alpha_H = f_H \, \sqrt{M_H}$
 (see \cite{neu,ahjoe}).
For the $W$-boson materializing to a   $D$ or $\bar{D}$ mesons,
 we obtain the bosonized current 
\begin{equation}
 \overline{q_L} \gamma^\alpha   Q_{c v}^{(\pm)} \;  \longrightarrow \;
    \fr{\alpha_H}{2} Tr\left[\xi^\dagger \gamma^\alpha L 
 H_{c q}^{(\pm)} \right]
  \; ,
\label{Jqc}
\end{equation}
where $v$ is the velocity of the heavy $c$ or  $\bar{c}$ quarks
($v=v_c$ or $v=v_{\bar{c}}$), and 
$H_{c q}^{(\pm)}$ is the corresponding field for the  $D_q$ or  $\bar{D_q}$ meson.

In order to calculate the matrix elements of the quark operators in 
(\ref{Lquark}) beyond the factorizable limit, we will use 
a model which incorporates emission of soft gluons modeled by a gluon
condensate. This will be performed within the 
HL$\chi$QM recently developed in \cite{ahjoe}.
See also \cite{chiqm,BHitE}.
 The Lagrangian for the HL$\chi$QM is 
\begin{equation}
{\mathcal L}_{\text{HL$\chi$QM}} =  {\mathcal L}_{\text{HQEFT}} +
  {\mathcal L}_{\chi\text{QM}}  +   {\mathcal L}_{\text{Int}} \; .
\label{totlag}
\end{equation}
The first term is given in equation (\ref{LHQEFT}).
The light quark sector is described by the Chiral Quark Model ($\chi$QM),
having a standard QCD term and a term describing interactions between
quarks and  (Goldstone) mesons: 
\begin{equation}
{\mathcal L}_{\chi\text{QM}} =  
\chibar \left[\gamma^\mu (i D_\mu     +  
\gamma_5  {\mathcal A}_{\mu})    -    m \right]\chi  \, + \, ...  \; ,
\label{chqmR}
\end{equation}
where the ellipses denote terms which are  irrelevant here.
Here $m$ is the SU(3) invariant constituent light quark mass, and
 $\chi$ is the flavour rotated quark fields given by
$\chi_L  =   \xi^\dagger q_L \; \; , \;  \chi_R  =   \xi q_R$, 
where $q^T  =  (u,d,s)$ are the light quark fields. The left- and
 right-handed
 projections $q_L$ and $q_R$ are transforming after $SU(3)_L$ and $SU(3)_R$
respectively. In (\ref{chqmR}) we have discarded terms involving the light
 current quark mass which is irrelevant in the present paper.
The covariant derivative $D_\mu$ in (\ref{chqmR}) is given as in (\ref{LS1})
and contains in addition  the soft gluon field forming the gluon condensates. The gluon 
condensate contributions are calculated by Feynman diagram techniques
as in \cite{ahjoe,EHP,ahjoeB,epb,BEF,pider}

The interaction between heavy meson fields and heavy quarks are
described by the following Lagrangian \cite{ahjoe}:
\begin{equation}
{\mathcal L}_{Int}  =   
 -   G_H \, \left[ \chibar_a \, \overline{H_a^{(\pm)}} 
\, Q^{(\pm)}_{v} \,
  +     \overline{Q_{v}^{(\pm)}} \, H_a^{(\pm)} \, \chi_a \right] \; ,
\label{Int}
\end{equation}
where  $G_F$ is a  coupling constant satisfying 
$G_H^2 = 2 m \rho/f_\pi^2$, $\rho$ being a hadronic parameter of order one.
In \cite{ahjoe} it was shown
how (\ref{LS1}) could be obtained from the HL$\chi$QM.
Performing this bosonization of the HL$\chi$QM,  one encounters divergent loop
integrals which will in general be quadratic-, linear- and
logarithmic divergent \cite{ahjoe}.
 Also, as in the
light sector \cite{BEF} the quadratic and logarithmic integrals are
related to the quark condensate and the gluon condensate respectively.

To calculate the factorizable contributions 
in (\ref{FactorizedP1}) and (\ref{FactorizedP3}) corresponding to Fig.~\ref{fig:BDfact}
within our framework, we
need the bosonized currents in (\ref{J(0)}) and (\ref{Jqc}), and in
addition
the bosonized currents involving an emision of a photon from the $B$-
or the $D$-meson.
For photon emission from the $B$-meson we have (for $v=v_b$)
\bea
&&\left(\overline{q_L}\,   \,\gamma^\alpha \, Q_{b v}^{(+)}\right)_{\gamma} 
\;   \longrightarrow\nonumber \\&&\qquad  
- \fr{G_H \, e}{32 \pi} \,F_{\mu\nu}
Tr\left[\xi^\dagger \gamma^\alpha  L \, H_{qb}^{(+)} \, Q_\xi 
\left( \sigma^{\mu\nu} \, - \,
 \fr{2 \pi f_\pi^2}{m^2 \, N_c}  \{\sigma^{\mu\nu},
 \gamma \cdot v \} \,
 \right)\right] \; ,
\label{1gamma}
\eea
where $F$ is the electromagnetic tensor and $Q_\xi = (\xi Q \xi^\dagger
 + \xi^\dagger Q \xi)/2$.
 For emision from the $D$-meson there is a
similar expression.

Bosonizing currents with one gluon emision from a coloured current
 operator, for instance to be used in the
   left part in Fig.~\ref{fig:Nonfact} b) 
 one obtains:
\bea
&&\left(\overline{q_L}\, t^a  \,\gamma^\alpha \, Q_{b v}^{(+)}\right)_{G} 
\;   \longrightarrow\nonumber \\&&\qquad  
- \fr{G_H \, g_s}{64 \pi} \,G_{\mu\nu}^a
Tr\left[\xi^\dagger \gamma^\alpha  L \, H_{b q}^{(+)}
\left( \sigma^{\mu\nu} \, - \,
 \fr{2 \pi f_\pi^2}{m^2 \, N_c}  \{\sigma^{\mu\nu},
 \gamma \cdot v \} \,
 \right)\right] \; ,
\label{1G}
\eea
where $G^a_{\mu \nu}$ is the octet gluon tensor, and
 $H_{bq}^{(+)}$ represents the heavy $\bar{B_q}$-meson fields.
Similarly the (heavy) $D$- and $\bar{D}$-mesons are represented by
$H_c^{(+)}$ and $H_{\bar{c}}^{(-)}$ corresponding to a heavy quark field
$Q_{v_c}^{(+)}$ and heavy anti-quark field $Q_{\bar{v}}^{(-)}$ respectively.
$v_c$ and $\bar{v}=v_{\bar{c}}$ are the velocities of the $c$ and
$\bar{c}$ quarks, respectively.
The symbol $\{\; , \; \}$  denotes the anti-commutator.

For one gluon and one photon emision from the $\overline{B_q}$-meson
appearing in  left part in Fig.~\ref{fig:Nonfact}~a) 
 one obtains:
\bea
\left(\overline{q_L}\, t^a  \,\gamma^\alpha \, Q_{v_b}^{(+)}\right)_{G\gamma} 
\;   \longrightarrow\nonumber \qquad  
G_H \, g_s \, e \, F_{\mu \nu} \,G_{\sigma \rho}^a
Tr\left[\xi^\dagger \gamma^\alpha  L \, H_{b q}^{(+)}
 R^{\mu \nu \sigma \rho} 
 \right] \; ,
\label{Ggam}
\eea
where the tensor $R$ contains products of Dirac matrices and
propagators
 (with momentum integrated out).
Multiplying the currents for each vertex, for instance those 
in eqs. (\ref{1G}) and (\ref{Ggam}),
and using   the prescription:
\begin{equation}
g_s^2 G_{\mu \nu}^a G_{\alpha \beta}^a  \; \rightarrow 4 \pi^2
 \gc \frac{1}{12} (g_{\mu \alpha} g_{\nu \beta} -  
g_{\mu \beta} g_{\nu \alpha} ) \, ,
\label{gluecond}
\end{equation}
we obtain the bosonized version for the operator $\widetilde{Q_1}$
in eqs. (\ref{QCol12}) and  (\ref{QCol34}) as the product of two
traces. (The expression may be simplified by using the Dirac algebra, but we do
not enter these details here).

\section{Amplitudes for $B \rightarrow \gamma \, D^*$ }

\begin{figure}[t]
\begin{center}
   \epsfig{file=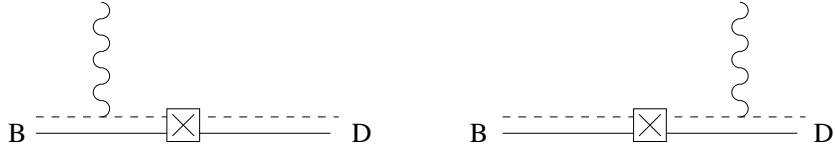,width=11cm}
\caption{Pole contributions to
$B \rightarrow \gamma \, D$ present for charged mesons. The combined
  full and dashed lines are the heavy mesons, and the wavy lines
  represent  photons.}
\label{fig:chiral1}
\end{center}
\end{figure} 


We restrict ourselves to processes where the $b$-quark decays, namely
the charged modes  $B^- \rightarrow \, \gamma D_{q}^{*-}$
and the neutral modes  $\overline{B_{q}^0} \rightarrow \,
\gamma \overline{D^{*0}} $, 
and  $\overline{B_{q}^0} \rightarrow \, \gamma D^{*0}$, for $q=d,s$. Processes
  where the  anti-$b$ quark decays proceed analogously.

Considering simple quark diagrams only,
 we observe that in terms of the Wolfenstein parameter
$\lambda$, the amplitudes for  $B^- \rightarrow \, \gamma D_{d}^{*-}$
and   $\overline{B_{d}^0} \rightarrow \,
\gamma \overline{D^{*0}}$ are ${\cal O}(\lambda^4)$ and small. In contrast, 
 the amplitude for $\overline{B_{d}^0} \rightarrow \, \gamma D^{*0}$ 
is  ${\cal O}(\lambda^2)$, and is KM non-suppressed compared to other
$b \rightarrow \gamma D^*$ modes..  
For $q=s$, all the amplitudes are  ${\cal O}(\lambda^3)$.
We will however see that strong interactions might make this simple
 picture more complicated.

For the charged decay(s) there are  {\em pole diagrams}, which are 
absent for the neutral decays. These are (within HQEFT) obtained by
the bosonized currents in (\ref{J(0)}) and (\ref{Jqc}) and the  photon
emission is obtained from (\ref{LS1}). Here the  $B^- \rightarrow  D_{s,d}^{*-}$
transitions are  proportional to the favorable coefficient 
$a_f = (a_{2,4} + a_{1,3}/N_c) \sim 1$,
while the non-factorizable contributions proportional to 
 $2 a_{1,3}$ due to coloured operators are relatively small.

\begin{figure}[t]
\begin{center}
   \epsfig{file=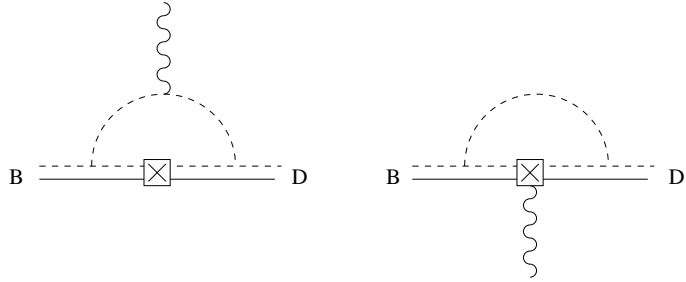,width=9cm}
\caption{Meson exchange diagrams. The combined
  full and dashed lines are the heavy mesons,and the single dashed
  lines represent light pseudoscalar mesons.}
\label{fig:mesonloop}
\end{center}
\end{figure}

There are  some {\em meson exchange} decay mechanism 
 contributions.
In the heavy quark
limit these are identical with chiral loop contributions.
These are shown in Fig.~\ref{fig:mesonloop}. For the process 
 $B^- \rightarrow \, \gamma D_{s,d}^{*-}$ there is an intermediate  
 $\overline{B_{d}^0} \rightarrow \, \overline{D^{*0}}$ transition
 accompanied with emission and re-absorpsion of $\pi^-$, or an 
 intermediate  
 $\overline{B_{s}^0} \rightarrow \, \overline{D^{*0}}$ transition
 accompanied with emission and re-absorpsion of $K^-$.
 For the process 
 $\overline{B_{d}^0} \rightarrow \,
\gamma \overline{D^{*0}} $ there is an intermediate 
 $B^- \rightarrow \,  D_{s,d}^{*-}$
   transition
 accompanied with emission and re-absorpsion of $\pi^+$. 
For  $\overline{B_{s}^0} \rightarrow \,
\gamma \overline{D^{*0}} $ there is an intermediate 
 $B^- \rightarrow \,  D_{s}^{*-}$  transition
 accompanied with emission and re-absorpsion of $K^+$.
Because the transitions  $B^- \rightarrow \,  D_{d,s}^{*-}$
are non-suppressed in the factorized limit, the decays
 $\overline{B_{d,s}^0} \rightarrow \,
\gamma \overline{D^{*0}}$ is semi-suppressed, having  meson exchange
amplitudes reducing to chiral loops in the HQEFT limits.
In this limit the meson exchange amplitudes are proportional to 
\bea
\chi(M) \; = \; \Big(\frac{g_A m_M}{4 \pi f_\pi}\Big)^2 
\ln(\frac{\Lambda_\chi^2}{m_M^2}) \; ,
\eea
for exchange of $M=K,\pi$ respectively. Here $g_A$ is the light
 ($M=K,\pi$) meson axial coupling to heavy mesons and $\Lambda_\chi
 \simeq $ 1 GeV.  Numerically, $\chi(K) \simeq 0.09$ and  $\chi(\pi)
 \simeq 0.02$, respectively. For the processes  
$\overline{B_{d,s}^0} \rightarrow \, \gamma D^{*0}$ there are only
 Zweig-forbidden
and $SU(3)_F$ violating neutral meson exchange which give small
 contributions.

For the processes 
$\overline{B_{d,s}^0} \rightarrow \, \gamma D^{*0}$ the amplitudes
$A^{(\pm)}$ are of the form
\bea
A^{(\pm)}_G \; = \; - \, \frac{e C_2}{2^8 \pi} \, G_H^2 \, \gc \,
  \left( Q_q^B \, Z^{(\pm)}_B + Q_q^D \,  Z^{(\pm)}_D \right) \; ,
\eea
 where $Q_q^B$ and $Q_q^D$ are the charges of the light quarks within
 the $B$ and $D$ mesons, respectively. For the processes 
$\overline{B_{d,s}^0} \rightarrow \, \gamma D^{*0}$ we have $Q_q^B=-1/3$ and $Q_q^D=2/3$.
 The quantities  $Z^{(\pm)}$ are of order one and given by
\bea
  Z^{(+)}_B \; = \;  (\frac{89 \pi}{288} -  \frac{13}{18} ) k \omega y 
            +  (\frac{7 \pi}{144} + \frac{5}{18}) k \omega^2 
+  (\frac{11}{18} - \frac{13 \pi}{96}) k 
   + (\frac{2}{3} - \frac{\pi}{18})\; ,
\eea
\bea
  Z^{(-)}_B \; = \; - \frac{5 \pi}{9} k \omega y + \frac{(\pi +2)}{9} k \omega^2
+ (\frac{\pi}{9}  - \frac{4}{3}) k - \frac{(\pi + 8)}{9}  \; , 
\eea
\bea
  Z^{(+)}_D \; = \; -(\frac{11 \pi}{288} + \frac{17}{36}) k \omega y
  + (\frac{1}{36}- \frac{53 \pi}{288}) k - (\frac{\pi}{64} +
  \frac{7}{72}) \omega y 
+  (\frac{41 \pi}{576} + \frac{23}{72}) \; ,
\eea
\bea
  Z^{(-)}_D \; = \; -(\frac{\pi}{3} + \frac{4}{9}) k + 
(\frac{2}{9} - \frac{\pi}{18})  \omega y +
  \frac{4}{3} k \omega y +  (\frac{2}{9} - \frac{\pi}{18}) \; , 
\eea
where the dimensionsless parameters $k, \omega$, and  $y$ are given by
\bea
k = \frac{2 \pi f_\pi^2}{N_c m^2} \; , \quad  \omega = v_b \cdot v_c \, = 
\, \frac{M_B^2 + M_D^2}{2 M_B M_D} \; , \quad y \, = \, \frac{M_B}{M_D} \; ,
\eea
where we for $M_D$ have used the mass of $D^*$. Using $m=230$ MeV,
$\rho = 1.1$, and $\gc^{1/4}$= 310 MeV, we obtain
\cite{JAMcDS} 
\bea
\label{BR}
 BR(\overline{B^0_d} \to \gamma  D^{*0}) \simeq 
 1.6 \times 10^{-6}  \; \; , \text{and} \quad
 BR(\overline{B^0_s} \to \gamma \, D^{*0} ) \simeq 8 \times 10^{-8}  \; \; .
\eea

For the  decays $\overline{B^0_{d,s}} \to \gamma  \overline{D^{*0}}$, there is a
delicate balance between different amplitudes, and it is hard to
conclude anything within our framework. The decays  
$B^- \to \gamma  D_{d,s}^-$, the factorizable contributions dominate,
and the amplitudes obtained from the diagrams in FIG.~2 alone are
\bea
A^{(\pm)}_F \; = \; - \, (C_4 + C_3/N_c) \,  \frac{e N_c G_H \alpha_H}{16
  \pi}  \, Y^{(\pm)} \; ,
\eea
where 
\bea
Y^{(+)}=  Q_q^B \,  + Q_q^D \,(1 + k - k \omega y)
\eea
which within our framework is roughly four times the pole
contribution. For the parity violating case, one has
\bea
Y^{(-)}=  - Q_q^B \,(1+ 2 k)   + Q_q^D \,(1 + k \omega y) \; .
\eea
To obtain a parity violating pole term, the intermediate heavy meson(s) must
have positive parity. We find that $B^- \to \gamma  D_{d}^-$ has a
branching ratio of order  $2 \times 10^{-7}$ within our framework.. 

\begin{table}
\begin{center}
\begin{tabular}{|c|c|c|c|c|}
\hline
Process & Pole & Factorized & Soft gluon & Meson exchange\\ \hline
$B^- \rightarrow \, \gamma D_{d}^{*-}$ &
 $a_f \; \lambda^4$ & $a_f \; \lambda^4$ & $2 a_3 \lambda^4 $ & 
$a_{nf} \; \lambda^4 \; \chi(\pi^-) $ \\ \hline
$B^- \rightarrow \, \gamma D_{s}^{*-}$ &
$a_f \; \lambda^3$ &  $a_f \; \lambda^3$ & $2 a_3 \; \lambda^3 $& 
$a_{nf} \; \lambda^3 \; \chi(K^-) $ \\ \hline
$\overline{B_{d}^0} \rightarrow \, \gamma \,  \overline{D^{*0}}$ &
-   &  $a_{nf} \; \lambda^4 $ &  $2 a_4 \, \lambda^4 $ & 
$ a_{f} \; \lambda^4 \; \chi(\pi^+) $  \\ \hline
$\overline{B_{s}^0} \rightarrow \, \gamma \,  \overline{D^{*0}}$ &
-   &  $a_{nf} \; \lambda^3$ & $2 a_4 \; \lambda^3$ & 
$a_{f} \; \lambda^3 \; \chi(K^+) $ \\ \hline
 $\overline{B_{d}^0} \rightarrow \, \gamma \, D^{*0}$ &
- &  $a_{nf} \; \lambda^2$ & $2 a_2 \; \lambda^2$ & Zweig-forb. \\ \hline
 $\overline{B_{s}^0} \rightarrow \, \gamma \, D^{*0}$ &
- &  $a_{nf} \; \lambda^3$ & $2 a_2 \; \lambda^3$ & Zweig-forb. \\ 
\hline
\end{tabular}
\caption{An overview of contributions to 
   processes of the type $B \rightarrow \, \gamma \, D^*$}
\end{center}
\end{table}

\vspace{2cm}

\section{ Conclusion}

We have considered six decay modes of the type 
$B \rightarrow \, \gamma D$ generated by three (main) mechanisms:

\begin{itemize}
 
\item 
 
a) Factorized contributions of pole and non-pole type. These might be
proportional to the favorable Wilson coefficient combination $a_f \sim
1$, or the non-favorable coefficient combination $a_{nf}$ of order
$10^{-2}$.

\item

 b) Meson
exchanges, that is, some intermediate $B \rightarrow D$ transition
accompanied with an emission and re-absorption of a pseodoscalar
boson($\pi$ or $K$).

\item 

 c) Emission of soft gluons,- modelled by a gluon condensate. 
In the HQEFT limits, the mechanisms b) and c) are (formally) $1/N_c$ suppressed.
If the corresponding factorizable amplitude is proportional to the favorable
coefficient $a_f$, this mechanism gives a non-important contribution
$\sim a_{1,3}$. On the other hand, if the factorizable contribution is
proportional to the non-favorable coefficient $a_{nf}$, the
non-factorizable soft emision amplitude contribution is proportional
to the favorable coefficient $2 a_{2,4}$, and gives a significant contribution.
Contributions to the various $B \rightarrow \gamma D*$ modes are
qualitatively summarized in Table I.

\end{itemize}

The present analysis is performed  within HQEFT, both for the $b$- and
the $c$-quark. Formally, the modes $B \rightarrow \gamma D*$ are
``heavy to heavy'', but for precise estimates our framework is 
unrealistic \cite{GriLe} because the
energy gap between the  $b$- and the $c$-quark are significantly
bigger than 1 Gev, which is the scale of HL$\chi$PT and the
HL$\chi$QM. Therefore large $1/m_Q$ corrections (especially $1/m_c$
corrections) must be expected. Phrased in another way,- damping form
factors are expected to be present, and most probably, our estimates
are overestimates. 
Alternative estimates, based on other frameworks, for instance
considering the charm quark as ``light'', should be performed.
Still, we expect that we have obtained {\em amplitudes}
 of the right order of magnitude, while the branching rations might
be an order of magnitude off. Still, our conclusion about the
importance of the non-factorizable gluon emision for 
$\overline{B_{d,s}^0} \rightarrow \, \gamma D^{*0}$ should hold.

\vspace{0.7cm}


{\acknowledgments}

JOE is supported in part by the Norwegian
 research council
 and  by the European Union RTN
network, Contract No. HPRN-CT-2002-00311  (EURIDICE).

\bibliographystyle{unsrt}

\end{document}